\documentclass[prl,showpacs,twocolumn,a4paper]{revtex4}
\usepackage{amsmath}
\usepackage{amsfonts}
\usepackage{amssymb}
\usepackage{graphicx}

\usepackage{color}
\usepackage{array,tabularx}

\begin{document}

\title{Size and density avalanche scaling near jamming}

\author{Roberto Ar\'evalo}
\author{Massimo Pica Ciamarra}\email[]{massimo.picaciamarra@spin.cnr.it}
\affiliation{
CNR--SPIN, Dipartimento di Scienze Fisiche,
Universit\`a di Napoli Federico II, I-80126, Napoli, Italy
}

\date{\today}

\begin{abstract}
The current microscopic picture of plasticity in amorphous assumes
local failure events to produce displacement fields complying with linear elasticity.
Indeed, the flow properties of nonaffine systems such as foams, emulsions and granular materials close to jamming, 
that produce a fluctuating displacement fields when failing, are still controversial.
Here we show, via a thorough numerical investigation of jammed materials, that nonaffinity induces
a critical scaling of the flow properties dictated by the distance to the jamming point.
We rationalize this critical behavior introducing a new universal jamming exponent and hyperscaling relations,
and use these results to describe the volume fraction dependence of the friction coefficient.
\end{abstract}
\maketitle

While in simple crystals particles experience the same amount of deformation
under a small, uniform applied stress, the disorder characterizing
amorphous materials leads to a fluctuating local deformation.
When this nonaffine contribution to the displacement field 
is of the same order of the affine displacement
itself, it strongly influence the 
elastic and rheological properties of the material and cannot be
handled via perturbative approaches~\cite{perturbative,Ohern2003,notaffine,Liu2010,vanHecke2010}.
In amorphous materials of technological interest
such as foams, emulsions, polymeric suspensions and granular materials,
this occurs close to jamming volume fraction $\phi_J$ marking the onset of
mechanical rigidity upon compression~\cite{Ohern2003}. 
Indeed, close to jamming nonaffinity effectively dominates the elastic response
being responsible, for instance, of an anomalous scaling of the shear to bulk modulus ratio, $\mu/k \propto \delta\phi^{1/2}$.
Nonaffinity also influences the rheological properties at a finite shear rate,
where its effects are known to depend on the energy dissipation mechanisms~\cite{Olsson2007,Tighe2010}.
Much less is known about the role of nonaffinity on the rheological properties
in the athermal quasistatic shear (AQS) limit, which is of particular interest as 
it allows for the identification of the microscopic plastic events,
and for the study of their properties and correlations.
In this limit plastic events result from
saddle node bifurcations~\cite{Maloney2006} that drive the irreversible rearrangement of a 
elementary unit of particles, generally known as a ``shear transformation zone'', STZ~\cite{STZ};
this elementary relaxation event might then trigger further rearrangements giving rise to avalanches,
whose spatial features, such as their fractal dimension, are controversial~\cite{Maloney2004,Maloney2006,Bailey2007}.
The triggering process is mediated by the elastic displacement field produced by STZs, experiments and simulations (e.g.~\cite{Maloney2006,BulatovArgon,Chikkadi2011})
frequently found to be alike that resulting from an Eshelby inclusion~\cite{Eshelby} in a linear elastic solid.
This suggests that in strongly non affine systems, where the displacement field produced by an elementary plastic event no longer resemble that produced by 
an Eshelby inclusion, the flow features might qualitatively and quantitatively change.
Recent AQS simulations~\cite{Heussinger2009,Heussinger2010} of harmonic disks in two dimensions
revealed that nonaffinity induces a critical behavior of some quantities characterizing the plastic flow, such as the average stress and energy drops, 
or the length of the elastic branches, which is dictated by the distance to the jamming point.
However, it is not known whereas this behavior is universal, 
as the role of the interaction potential and that of the dimensionality have not been explored,
and the critical exponents have not been theoretically rationalized. Similarly,
the effect of the increasing nonaffinity on the size scaling of the flow properties,
that reveal the geometrical features of the avalanches, is unknown.

Here we investigate the role of nonaffinity in the flow properties of amorphous systems via
AQS shear simulations of harmonic and Hertzian particles, in both
two and three dimensions, as a function of the distance
to the jamming threshold, $\delta \phi$, and of the system size, $N$.
We show that the macroscopic flow properties result qualitatively unaffected by the degree of nonaffinity, 
as the dynamics exhibits the same size regardless of the distance to the jamming point.
Nonaffinity controls the critical scaling of the dynamics with $\delta \phi$; we
rationalize this scaling showing that the exponent characterizing
the critical behavior of the length of the elastic branches is universal, 
and introducing hyperscaling relations involving the critical exponents
and the interaction potential. 
Finally, we use these results to infer the behavior of the friction coefficient at the jamming threshold.



We consider 50:50 binary mixtures of $N$ particles with diameters $\sigma_1=1$ and $\sigma_2=1/1.4$ 
and unit mass, in $d = 2$ and in $d = 3$ dimensions. Particles interact via 
a finite range repulsive potential, $V(r) = \varepsilon \left(\frac{\sigma-r}{\sigma_1}\right)^\alpha$ for $r < \sigma$, 
$V(r) = 0$ for $r > \sigma$, corresponding to an harmonic ($\alpha = 2$) or to an Hertzian potential ($\alpha = 3/2$), 
with $\sigma$ average diameter of the interacting particles. 
$\varepsilon$ and $\sigma_1$ are our units of energy and length, respectively.
We investigate different values of the volume fraction, considering systems 
with $N = 400,800,1600,3200$ for both harmonic and Hertzian particles, in both 2d and 3d,
as well as, for harmonic particles in 2d, $N=6400$ and $12800$.
These systems have been previously throughly investigated
in the jamming context~\cite{Ohern2003}, and that the pressure, the bulk and the shear 
modulus are known to scale with the overcompression as
$p \propto \delta\phi^{\alpha-1}$, $k \propto \delta\phi^{\alpha-2}$, and as $\mu \propto \delta\phi^{\alpha-3/2}$.
Here we deform them via athermal quasistatic shear (AQS) simulations in which the strain
is increased by small steps $\delta\gamma=10^{-5}$, and the energy of the system is minimized
via the conjugate--gradient protocol after each strain increment~\cite{lammps}. Results are robust
with respect to a factor $10$ change in $\delta \gamma$. 
In the steady state regime of the AQS dynamics ($\gamma > 1$), we have identified 
all plastic events, and measured the yield stress $\sigma_Y$, the stress $\Delta \sigma$
and the energy $\Delta E$ drops, the strain interval between
successive events, $\Delta \gamma$, and the instantaneous shear modulus, $\mu$. 
For each value of $N$ and $\delta \phi$, we have recorded from $10^3$ to $10^4$ plastic events.

\begin{figure}
\includegraphics*[scale=0.35]{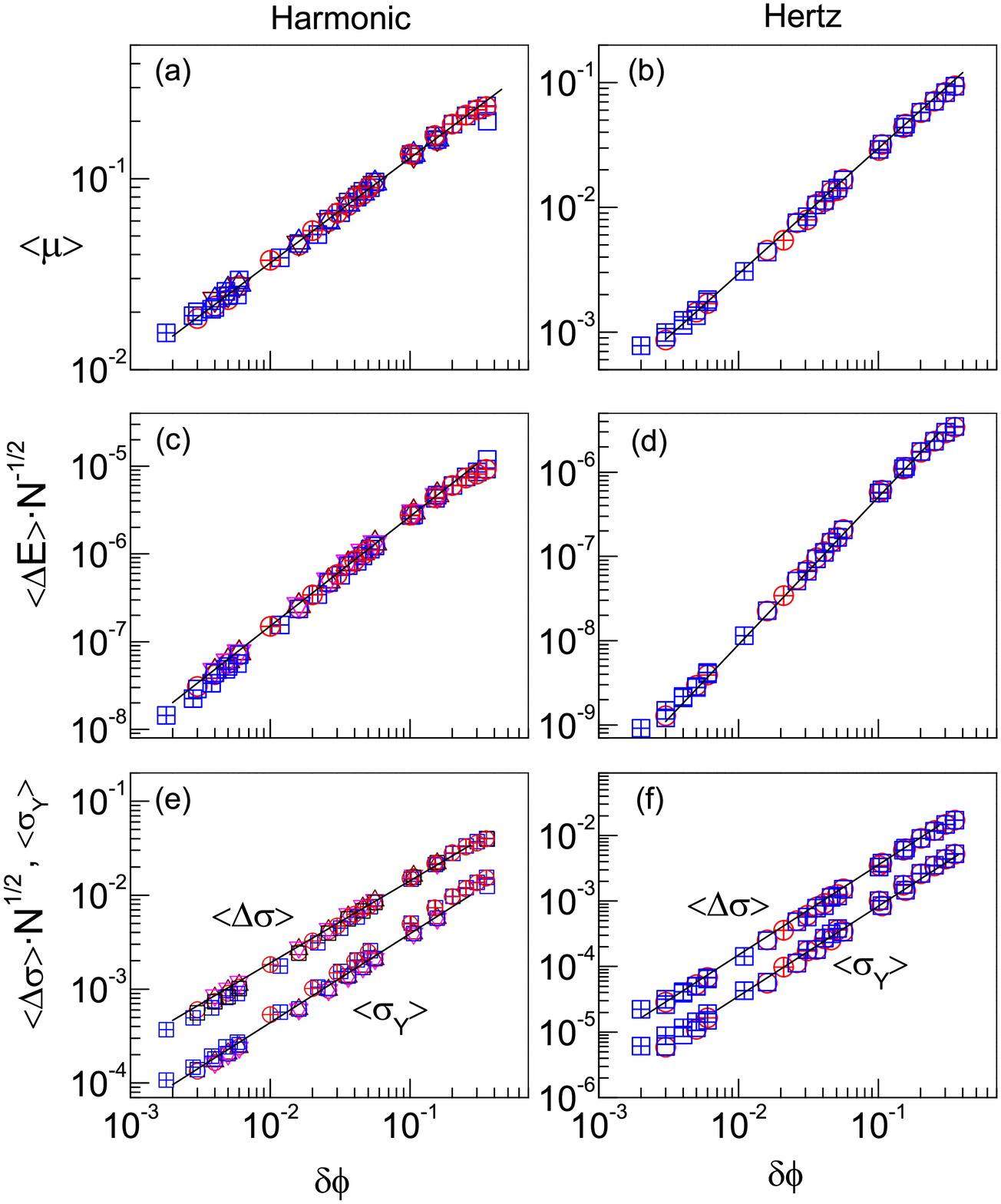}
\caption{Scaling of the shear modulus (panels a and b), of the energy drops (panels c and d),
of the yield stress and of the stress drops (panels e and f), for harmonic (left column)
and Hertzian particles (right column). Symbols refer to $N = 1600$ (squares), $3200$ (circles),
$6400$ (up triangles) and $12800$ (down triangles). A superimposed cross distinguishes the $3d$ data.
The data collapse occurs with no vertical shifts. Lines are power--law with exponents summarized in
Table~\ref{thetable}.
}
\label{fig:averages}
\end{figure}

\begin{figure}
\includegraphics*[scale=0.33]{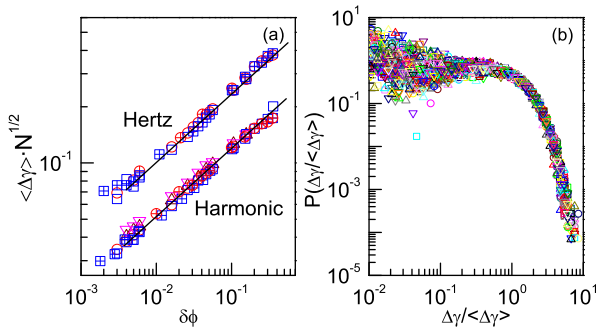}
\caption{(a) Universal scaling of the average length of the elastic branches. Symbols are as in Fig.~\ref{fig:averages}.
Data for Hertzian particles are shifted by a factor $2$ for clarity. (b) Scaling of the $\Delta\gamma$ distribution. Data
from $100$ simulations obtained changing $N$, $\delta \phi$ and interaction potential collapse on the same master curve.}
\label{fig:gamma}
\end{figure}

Our results support the presence of scaling relations of the form
$\langle X\rangle \propto N^{\omega_X}\delta\phi^{\nu_X}$, where $X$ is one of the investigated
quantities, and $\omega_X$ and $\nu_X$ its size and density scaling exponents, respectively.
Data supporting the validity of these relations are illustrated in Fig.~\ref{fig:averages}
and in Fig.~\ref{fig:gamma}a. For sake of clarity, we approximate the exponents with simple integer fractions to which the measured
values agree within $\approx 0.02$, and summarize their values in Table~\ref{thetable}. 

The size scaling is of interest as a signature of the
microscopic features of the plastic events and of their correlations~\cite{Maloney2006,Bailey2007,LemaitreCaroli},
that build up through the elastic displacement field induced by the STZs.
For instance, if plastic avalanches have a fractal dimension $D$, then $\omega_{\Delta E} = D/d$ in
$d$ spatial dimensions~\cite{Maloney2006,Bailey2007}.
Previous works have investigated the size
scaling via AQS simulations of a variety of different systems, 
confirming the intensive character of the shear stress and of the shear modulus, 
$\omega_{\sigma_Y} = \omega_\mu = 0$, and revealing a size dependence of other quantities
associated to the plastic events. 
The energy--stress ($\Delta E \propto N \Delta \sigma$) and the stress--strain ($\Delta \sigma \propto \mu \Delta \gamma)$
dependence lead to the relations $\omega_{\Delta E} - \omega_{\Delta \sigma} = 1$,
and $\omega_{\Delta \gamma} = \omega_{\Delta \sigma}$. 
These are always verified in the literature,
most works reporting values compatible with $\omega_{\Delta E} = 1/2$
regardless of the dimensionality~\cite{Maloney2004,Maloney2006,LemaitreCaroli,Tanguy2006,Bailey2007,Heussinger2010},
even though values compatible with $\omega_{\Delta E} = 1/3$ have also been reported~\cite{Lerner2009}.
Our data of Fig.s~\ref{fig:averages} and~\ref{fig:gamma}a show that both away form the jamming transition, where the system's response is affine,
as well as close to the transition, where the response is highly non--affine, 
the exponents assume values compatible with $\omega_{\Delta E} = 1/2$, $\omega_{\Delta \sigma} = -1/2$ and $\omega_{\Delta \gamma} = -1/2$.
These values do not depend neither on the interaction potential, nor on the dimensionality.
These results clarify that the macroscopic plastic flow properties are unaffected by the nonaffine local response of the system.

We now focus on the scaling with respect to the distance from the jamming point, $\delta \phi$. 
For each investigated quantity and system size, we have performed a power law fit 
to extract the critical exponent and the critical volume fraction, $\phi_j$. The resulting values of $\phi_j$
are compatible within $10^{-3}$, consistently with the observation of a weak system size dependence
under shear~\cite{Vagberg2011}, and have average values $\phi_j=0.843$ in $2d$, and $\phi_j=0.645$ in $3d$.
These values are in good agreement with those reported for sheared systems~\cite{Olsson2011,Heussinger2010}, 
and we have used them to determine $\delta \phi = \phi-\phi_j$.
Fig.~\ref{fig:averages} summarizes our results for the scaling of the shear modulus, of the average energy drops,
of the average stress drops and of the yield stress.
The data collapse and the numerical fits clarify that the critical exponents 
characterizing the $\delta \phi$ scaling of these quantities depend on the interaction potential, not on the dimensionality. 
On the contrary, neither the interaction potential nor the dimensionality affect the scaling of 
the length of the elastic branches, we find to be characterized
by a universal scaling exponent, $\nu_{\Delta \gamma} \approx 3/8$, as illustrated in Fig.~\ref{fig:gamma}a.

\begin{figure}
\includegraphics*[scale=0.33]{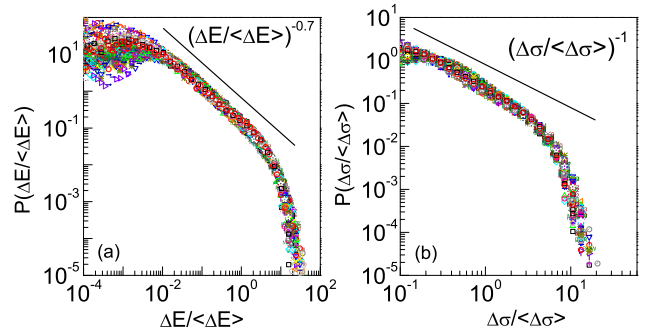}
\caption{Scaling of the distribution of the energy drops (a) and of the stress drops (b).
As in Fig.~\ref{fig:gamma}b, data from 100 simulations referring to different $N$, $\delta \phi$ and $\alpha$ collapse on the same master curve.}
\label{fig:distributions}
\end{figure}

The scaling relations we have illustrated referring to the average values
do actually work for the whole distributions, 
that scale as $P(X) = \langle X \rangle g_X(X/\langle X \rangle)$. Here
$g_X$ is a scaling function that depends on the considered quantity, but not
on the dimensionality, the interaction potential or the volume fraction.
Fig.~\ref{fig:gamma}b illustrates this scaling for the length of the elastic branches 
reporting the collapse of $100$ datasets referring to different system sizes,
densities and dimensionalities. 
Similarly, Fig.~\ref{fig:distributions} shows the collapse of the distributions of the energy and of the stress drops.
These two distributions have a roughly power law initial decay, which is followed by an exponential cut-off. 
In the case of the stress drop, the initial power law exponent is
$\approx -1$, as found~\cite{Heussinger2010} for $d=2$ harmonic particles.
For the energy drop we find an initial power law exponent $\approx -0.7$,
in agreement with earlier results~\cite{Tewari1999,Maloney2004} for $d=2$ harmonic particles.
As an aside, we note that repulsive systems sheared via a spring mechanism are common experimental and numerical models for earthquakes~\cite{earthquakes}, 
the particles representing the fault gouge in between the tectonic plates. These studies have found
a different power law exponent for the energy drop distribution, $\approx -1.7$, 
in agreement with the well known Gutenberg--Richter law.
This suggests that in purely repulsive systems inertia affects the scaling exponents and the scaling distributions, 
as recently observed in LJ systems~\cite{salerno}.

We now show the existence of hyperscaling relations between the critical exponents.
We begin by noticing that there are no correlations between the measured quantities (not shown), so that
$\langle\Delta\sigma\rangle=\langle\mu\Delta\gamma\rangle=\langle\mu\rangle\langle\Delta\gamma\rangle$. Inserting 
the corresponding scaling relations we recover the known relation $\omega_{\Delta \sigma}=\omega_{\Delta \tau}$ for the size scaling, 
and find a relation for the density scaling,
\begin{eqnarray}
\nu_{\Delta\sigma}=\nu_{\mu}+\nu_{\Delta\gamma}. 
\label{eq:relation1}
\end{eqnarray}
We then consider that, due to the quadratic relation between energy and strain, 
the energy released in a plastic event is $\Delta E=
\frac{1}{2\rho}\frac{N}{\mu}\left[2\sigma_Y\Delta\sigma-\Delta\sigma^2\right]$.
Given the scaling of the two term in brackets, the condition $\Delta E > 0$
can be expressed as $N^{\beta}\delta\phi^{\nu_{\sigma_y}-\nu_{\Delta\sigma}} > 1$,
and is always satisfied if
\begin{eqnarray}
\nu_{\sigma_Y}=\nu_{\Delta\sigma}.
\label{eq:relation2}
\end{eqnarray}
Conversely, the condition could be violated at small or at large $\delta \phi$.
Given Eq.~\ref{eq:relation2}, the above equation for the energy drop leads to the 
already known relation for the size scaling, $\omega_{\Delta E}-\omega_{\Delta \sigma}=1$,
and to a new relation for the density scaling,
\begin{eqnarray}
\nu_{\Delta E} = -\nu_{\mu} + 2\nu_{\Delta\sigma} = \nu_{\mu}+2\nu_{\Delta\gamma}.
\label{eq:relation3}
\end{eqnarray}
The validity of these relations, that can be also derived from a simple dimensional analysis, is easily verified from 
the results summarized in Table~\ref{thetable}.
As concern the dependence on the overcompression $\delta \phi$, we have investigated $5$ exponents,
and derived three hyperscaling relations. The only independent exponents are that of the shear modulus,
which is known to be fixed by the interaction potential, $\nu_{\mu} = \alpha-3/2$~\cite{Ohern2003}, 
and that of the length of the elastic branches, we have found to be universal, $\nu_{\Delta\gamma} = 3/8$.

\begin{center}
\begin{table}
\begin{tabular}{c c c c }
\hline
$\langle X \rangle \propto N^{\omega_X}\delta\phi^{\nu_X}$~~  &~Harmonic~ &~~~~Hertz~~~~& Relations \\
\hline\hline
$\omega_{\Delta E}$             & $1/2$   & $1/2$  &   \\
$\omega_{\Delta \sigma}$              & $-1/2$  & $-1/2$ &   $\omega_{\Delta \sigma} = \omega_{\Delta E} - 1$\\
$\omega_{\Delta \gamma}$               & $-1/2$  & $-1/2$  &  $\omega_{\Delta \gamma} = \omega_{\Delta E} - 1$ \\
\hline
$\nu_{\mu}$          & $1/2$   & $1$    & $\nu_{\mu} = \alpha-3/2$   \\
$\nu_{\Delta\sigma}$ & $7/8$   & $11/8$ & $\nu_{\Delta\sigma} = \nu_{\mu}+\nu_{\Delta\gamma}$  \\
$\nu_{\sigma_Y}$     & $7/8$   & $11/8$ & $\nu_{\sigma_Y} = \nu_{\mu}+\nu_{\Delta\gamma}$  \\
$\nu_{\Delta E}$     & $5/4$   & $7/4$  & $\nu_{\Delta E} = \nu_{\mu}+2\nu_{\Delta\gamma}$  \\
$\nu_{\Delta\gamma}$ & $3/8$   & $3/8$  &  
\end{tabular}
\caption{Values of the exponents describing the scaling as a function of the 
system size, $\omega$, and of the overcompression, $\nu$. The exponents do not depend
on the dimensionality. Our numerical fits yield values consistent with the reported integer
fractions, that satisfy the indicated relations.}
\label{thetable}
\end{table}
\end{center}

\begin{figure}
\includegraphics*[scale=0.33]{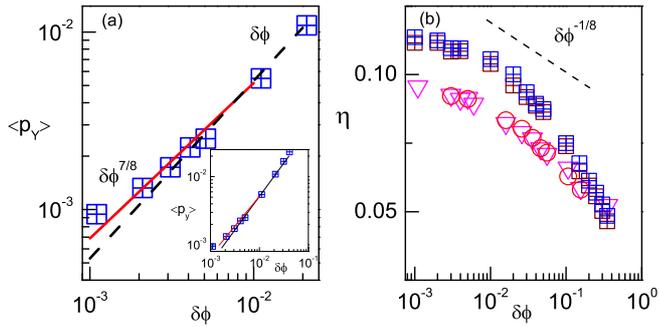}
\caption{ (a) Crossover in pressure dependence on the overcompression $\delta \phi$, for 3d harmonic spheres. 
The inset and the main panel show the whole dependence and a zoom on the crossover region, respectively.
(b) Scaling of the friction coefficient for all considered system sizes and potentials. Symbols are as in Fig.~\ref{fig:averages}.
}
\label{fig:friction}
\end{figure}

We finally consider the behavior of the pressure along the yield stress line, $p_Y$. First, we notice that
a power law fit of our pressure data in the investigated volume fraction range gives
a critical exponent that is slightly larger than that observed at zero applied shear stress, ${\alpha-1}$,
consistently with previous results for 2d harmonic disks~\cite{Heussinger2010,Olsson2011}.
However, the fit gives a critical volume fraction that is smaller and not compatible with that resulting from
the other power laws fit. We interpret this as a signature of the fact that the pressure does not behave as a simple
power law, being affected by both the compression and the shear stress.
Indeed, when plotted as a function of $\delta\phi$, as in Fig.~\ref{fig:friction}a for 3d harmonic spheres,
the pressure exhibits two regimes. At high compressions, the pressure roughly scales 
as $\delta\phi^{\alpha-1}$, as in the case of zero applied shear stress, while at small compressions
it scales with a smaller exponent, $q$. Simulations with a smaller value of $\delta \phi$, which are difficult
to obtain due to the high computational cost of AQS simulations close to jamming, are needed to estimate $q$ with confidence.
Nevertheless, dimensional analysis suggests $q = \nu_{\sigma_Y} = \alpha-9/8$, which is a value compatible with our data.
The behavior of the pressure leads to a crossover in the effective friction coefficient $\eta \propto \sigma_Y/p_Y$.
This is expected to approach a constant value close to jamming, as in previous numerical~\cite{friction_numerical}
and experimental works~\cite{friction_experimental}, as $q = \nu_{\sigma_Y}$, and to scale as
$\delta \phi^{-1/8}$ away from the transition, regardless of the dimensionality and of the interaction potential.
Fig.~\ref{fig:friction}b shows that these predictions are verified by our numerical data.

We have shown that the size scaling of the plastic flow features of amorphous
materials is surprisingly unaffected by the distance to the jamming point, and
therefore by degree of nonaffinity. This suggests that,
contrary to the common belief, Eshelby like displacement fields might not play a fundamental role in the plastic
flow, but more work is needed in this direction. 
Nonaffinity quantitaively influences the flow leading
to a critical scaling of the dynamics with the distance to the jamming thresold,
with exponents not depending on the dimensionality. 
We have rationalized this critial behavior introducing 
hyperscaling relations between the exponents, and a new universal jamming exponent.
This universality suggests a critical behavior of the energy landscape itself.
\begin{acknowledgments}
We thank R. Pastore for discussions, and MIUR-FIRB RBFR081IUK for financial support.
\end{acknowledgments}

\end{document}